\documentclass[conference]{IEEEtran}

 \usepackage[table]{xcolor}
\usepackage{url}
\usepackage{amsmath} 
\usepackage{graphicx} 
\usepackage{cite} 
\usepackage{color} 
\usepackage{soul} 
\usepackage[]{todonotes} 
\usepackage{bigstrut} 
\usepackage{listings} 
\usepackage{balance} 
\usepackage{algpseudocode} 
\usepackage{algorithm} 
\usepackage{algorithmicx} 
\usepackage{multicol}

\usepackage{amsthm} 
\usepackage{multirow} 
\usepackage{rotating} 
\usepackage{verbatim}
\usepackage{rotating} 
\usepackage{framed}
\usepackage{alltt}
\usepackage{graphicx}
\usepackage{caption}

\usepackage{lipsum}

\usepackage{color}
\usepackage{subfigure}
\usepackage{cite}

\newcommand{\mehdi}[1]{\textcolor{red}{{\it [Mehdi says: #1]}}}

\newcommand\totalProjects{82,447} 
\newcommand\totalProjectsWithTests{14,153} 
\newcommand\usedFrameworks{93}

\newcommand\totalLanguages{48}
\newcommand\totalFrameworks{251} 

\newcommand\totalPatternProjects{3,363} 
\newcommand\totalAllPatternProjects{11} 

\begin{document}


\title{A Large-Scale Study on the Usage of Testing Patterns that Address Maintainability Attributes \\\vspace{6pt}\Large{{Patterns for Ease of Modification, Diagnoses, and Comprehension}}}

\author{
    \IEEEauthorblockN{Danielle Gonzalez \IEEEauthorrefmark{1}, Joanna C.S. Santos\IEEEauthorrefmark{1}, Andrew Popovich\IEEEauthorrefmark{1}, Mehdi Mirakhorli\IEEEauthorrefmark{1}, Mei Nagappan\IEEEauthorrefmark{2}}
    \IEEEauthorblockA{\IEEEauthorrefmark{1}Software Engineering Department, Rochester Institute of Technology, USA}
    \IEEEauthorblockA{\IEEEauthorrefmark{2}David R. Cheriton School of Computer Science, University of Waterloo, Canada
\\\{dng2551,jds5109,ajp7560, mxmvse\}@rit.edu, mei.nagappan@uwaterloo.ca}
}
\maketitle
\begin{abstract}
Test case maintainability is an important concern, especially in open source and distributed development environments where projects typically have high contributor turn-over with varying backgrounds and experience, and where code ownership changes often. 
Similar to design patterns, patterns for unit testing promote maintainability quality attributes such as \textit{ease of diagnoses}, \textit{modifiability}, and \textit{comprehension}. In this paper, we report the results of a large-scale study on the usage of four xUnit testing patterns which can be used to satisfy these maintainability attributes.  This is a first-of-its-kind study which developed automated techniques to investigate these issues across 82,447 open source projects, and the findings provide more insight into testing practices in open source projects. Our results indicate that only 17\% of projects had test cases, and from the 251 testing frameworks we studied, 93 of them were being used. We found 24\% of projects with test files implemented patterns that could help with maintainability, while the remaining did not use these patterns. Multiple qualitative analyses indicate that usage of patterns was an ad-hoc decision by individual developers, rather than motivated by the characteristics of the project, and that developers sometimes used alternative techniques to address maintainability concerns.

\end{abstract}


\IEEEpeerreviewmaketitle

\section{Introduction}
The \textit{unit test} serves as an important facet of software testing because it allows individual ``units'' of code to be tested in isolation. With the development of automated testing frameworks, such as the \textit{xUnit} collection, writing and executing unit tests has become more convenient. 
However, writing quality unit tests is a non-trivial~\cite{perscheid2012test,BellerGPZ15} task. Similar to production code, unit tests often need to be read and understood by different people. The moment a developer writes a unit test, it becomes legacy code that needs to be maintained and evolve with changes in production code.  

Particularly, writing quality unit tests that encourage \textit{maintainability} and \textit{understandability} is a very important consideration for open source projects and the distributed development environment in general, where projects have high contributor-turn over, contributors have varying backgrounds and experience, and code ownership changes often~\cite{Samoladas:2004:OSS:1022594.1022598}.  

Existing research and best practices for unit testing focus on the quality of the test's design in terms of optimizing the coverage metric or defects detected. However, recent studies~\cite{Daka:2015:MRI:2786805.2786838, Fraser:2013:AWT:2483760.2483774,bavota2012empirical,Palomba:2016:ATC:2931037.2931057} emphasize that the maintainability and readability attributes of the unit tests directly affect the number of defects detected in the production code. Furthermore, achieving high code coverage requires \textit{evolving} and \textit{maintaining} a growing number of unit tests. Like source code, poorly organized or hard-to-read test code makes test maintenance and modification difficult, impacting defect identification effectiveness. 
In ``xUnit Test Patterns: Refactoring Test Code'' George Meszaro~\cite{Meszaros07} presents a set of automated unit testing patterns, and promotes the idea of applying patterns to test code as a means of mitigating this risk, similar to how design patterns are applied to source code.

Previous studies~\cite{kochhar2013empirical,bavota2012empirical,Neukirchen2007,greiler2013automated} have looked at test case quality in open source and industry, studying the effects of test smells and bad practices, and detecting quality problems in test suites. However, these and other testing-related works ~\cite{zaidman2008mining,lam2015parameterized} are limited to a small number of projects and have not taken into account use of testing frameworks and patterns that can be used to address maintainability attributes.  

In this paper, we aim to assess the satisfaction of test maintainability quality attributes in open source projects by studying the use of automated unit testing frameworks and the adoption of four of Meszaros' testing patterns within a large number of projects.

First, we conducted a large-scale empirical study to measure the application of software testing in the open source community, providing insights into whether open source projects contain unit tests and which testing frameworks are more common. Second, to evaluate the maintainability characteristics of unit tests written in the open source community, we have defined three quality attributes: \textit{Ease of Modification}, \textit{Ease of Diagnoses}, and \textit{Ease of Comprehension}. To assess the maintainability characteristics of the tests with regard to these quality attributes, we detect the adoption of four unit testing patterns that, when applied to unit tests within open source projects, help satisfy these attributes. Our novel approach includes a data set of \totalProjects{} open source projects written in 48 languages, and our approach for identifying pattern use is language-agnostic, increasing the generalization of our results to open source development as a whole.
We also conduct a qualitative study to reveal project factors which may have influenced a developer's decision whether or not to apply patterns to their tests. 

More precisely, the following motivating question and sub-questions have been explored and answered in this study: \textbf{Motivating Question:}\textit{ Do Test Cases in Open Source Projects Implement Unit Testing Best Practices to Satisfy Test Maintainability Quality Attributes?}
To accurately answer this question, we investigate the following sub-questions:

\begin{itemize}
\item \textbf{RQ1:} What percentage of projects with tests applied patterns that address \textit{Ease of Modification}?
\item  \textbf{RQ2:} What percentage of projects with tests applied patterns that address \textit{Ease of Diagnoses}?
\item  \textbf{RQ3:}  Do automatically generated test plans and checklists focused on testing breaking points improve the security and reliability of a software product?\item  \textbf{RQ3:} 
\end{itemize}

Despite investigating the use of only four testing patterns, our novel approach to measuring test quality will highlight important facts about the quality of test cases present in open source projects.

The reminder of this paper is organized as follows. A description of related works is provided in section~\ref{sec:related}, and section~\ref{sec:background} defines important terms used throughout the paper.
Section~\ref{sec:method} describes in detail the methodology for answering our three research questions. Results are presented in section~\ref{sec:results}. Qualitative analyses and discussion of our pattern results is presented in section~\ref{sec:patternanalysis}. Threats to validity are acknowledged in section~\ref{sec:threats}, and we end with our conclusions and proposed future work.

\section{Related Work} 
\label{sec:related}
In this section we identify relevant work studying unit testing in open source software, maintainability and quality of unit test cases, and the relationship between production and test code. 

The studies most closely related to the work presented in this paper were presented by Kochar et. al. This group was the first to conduct a large-scale study on the state of testing in open source software~\cite{kochhar2013empirical} using 20,817 projects mined from Github. They studied the relationship between the presence and volume of test cases and relevant software project demographics such as the number of contributing developers, the project size in lines of code (LOC), the number of bugs and bug reporters. They also measured the relationship between programming languages and number of test cases in a project.  
A year later, they conducted a quality-focused study ~\cite{kochhar2014empirical} using code coverage as a measure of test adequacy in 300 open source projects.

Other small-scale studies have evaluated the quality or effectiveness of unit tests in software projects. Badri et. al's used two open source Java projects to study levels of testing effort to measure the effectiveness of their novel quality assurance metric ~\cite{Badri2015529}. Bavota et. al~\cite{bavota2012empirical,bavota_are_2015} performed two empirical analyses to study the presence of poorly designed tests within industrial and open source Java projects using the JUnit framework, and the effects these bad practices have on comprehension during maintenance phases. Evaluations of the effectiveness of automated unit test generation tools show the importance of the usability and readability quality attribute, such as a study by Rojas et. al~\cite{rojas2015automated} on automatic unit tests generated by\textit{Evosuite}. To improve understandability of test cases Panichella et. al created a summarization technique called \textit{Test Describer}.~\cite{Panichella:2016:ITC:2884781.2884847}. Athanasiou et. al~\cite{athanasiou2014test} created a model to measure test code quality based on attributes such as maintainability found to positively correlate with issue handling.

Some works have taken an approach opposite to the one presented in this paper to measure the quality of unit tests by measuring the presence of `test smells' and bad practices within projects. These works uses the `test smells' defined by Van Deursen et. al~\cite{van2001refactoring} and broadened by Meszaros~\cite{Meszaros07}. 
Van Rompaey and Demeyer~\cite{van2007detection} defined a metric-based approach to detect two test smells, and later Brugelmans and Rompaey~\cite{breugelmans2008testq} developed \textit{TestQ} to identify test smells in test code. To reduce erosion of test maintainability over time due to fixture refactoring, Greiler et. al~\cite{greiler2013automated}, developed \textit{TestHound}, which detects test smells. 
Neukirchen et. al~\cite{Neukirchen2007} have used the presence of test smells to identify maintenance and reusability problems in test suites which use TCCN-3.

A number of works have been published which study the relationships between production code and unit tests. Using graphs obtained from static and dynamic analysis, Tahir and MacDonell~\cite{tahir2015combining} applied centrality measures to quantify the distribution of unit tests across five open source projects in order to identify the parts of those projects where the most testing effort was focused.
Zaidman et. al~\cite{zaidman2008mining} use coverage and size metrics along with version control metadata to explore how tests are created and evolve alongside production code at the commit level on two open source projects.
Van Rompaey and Demeyer~\cite{van2009establishing} have established conventions for improving the traceability of test code such as naming conventions, static call graphs, and lexical analysis. These methods have been used in other papers to ~\cite{tahir2015combining} establish heuristics for identifying tests in source code.

In this study we expand the scope of existing explorations of testing in open source by increasing the number of projects and languages being examined. The novel contributions presented in this paper are the use of three quality attributes, measured via pattern and test framework adoption in projects, as a measure of unit test maintainability in the open source community.

\section{Background and Definitions}\label{sec:background}
We begin by defining the important concepts and terms required to understand our work. In this section we define the metrics, three maintenance quality attributes for unit tests, and the testing patterns used throughout the paper.

\subsection{Project Metrics}
\label{sec:metrics}
\subsubsection*{Testing Ratio} 

To better understand the quantity of test code present in open source projects, we used the \textit{TPRatio} metric, defined as the Production Code to Unit Test Code Ratio: 

\[ TPRatio = \dfrac{Unit Test LOC}{Production LOC} \]

TPRatio measures the ratio of unit test lines of code to production (non-test) lines of code. Other works have used code coverage to measure test volume in projects~\cite{902502}, but we chose the TPRatio metric as a reasonable alternative because we do not have a reliable means of collecting coverage data for all the languages in our dataset. Ward Cunningham provides a discussion of this metric on his website~\cite{cunningham_2012} which mentions it as a `quick check' before measuring coverage. This metric has also been used in a previous study by Kochar et. al~\cite{kochhar2013empirical}.

\textit{Project Size}
\label{sec:projsize}
Project size within our dataset varies widely, which can impact how certain results, such as TPRatio, are presented. To prevent this, we divide projects in our dataset into four sizes when presenting results: 
\begin{itemize}
\item{Very Small: Project Size $<$ 1 kLOC}
\item{Small: 1 kLOC $\leq$ Project Size $<$ 10 kLOC }
\item{Medium: 10 kLOC $\leq$  Project Size $<$ 100 kLOC}
\item{Large: Project Size $\geq$ 100 kLOC}
\end{itemize}

\subsection{Unit Test Quality Attributes}
\label{sec:qualityattribs}
To evaluate the maintainability characteristics of unit tests written in the open source community, we have defined the following quality attributes, derived from literature ~\cite{Meszaros07,bavota2012empirical,van2001refactoring} that can impact the quality of unit tests. These attributes, summarized below, were used as driving requirements to identify patterns that improve the quality of unit tests. 

\begin{itemize}
\item \noindent\underline{QA\#1 - Ease of Modification}: The ability of unit tests in a system to undergo changes with a degree of ease.
\item \noindent\underline{QA\#2 - Ease of Diagnoses}: The measure of difficulty when attempting to track and inspect a test failure.
\item \noindent\underline{QA\#3 - Ease of Comprehension}:
The degree of understandability of a unit test being read by a developer not originally involved in its writing.
\end{itemize}

\subsection{Testing Patterns}
`Test smells', introduced by Van Deursen et. al~\cite{van2001refactoring}, negatively affect the maintainability of unit tests~\cite{bavota_are_2015,bavota2012empirical}. Guidelines for refactoring test code to remove these smells have been proposed~\cite{fowler1999refactoring,van2001refactoring} but in order to \textit{prevent} smells and promote high quality tests, Meszaros has detailed 87 goals, principles, and patterns for designing, writing and organizing tests using the \textit{xUnit} family of unit testing frameworks~\cite{Meszaros07,meszaros_test_2003}
Therefore, in order to evaluate the maintainability of unit tests in open source projects, we chose to measure the adoption of some of these patterns. We developed 3 criteria when choosing patterns; first, a pattern must address at least one of the quality attributes defined in section~\ref{sec:qualityattribs}. Next, a pattern must be optional for a developer to implement, i.e. not enforced by the framework. This will help us to understand the decisions open source developers make. Finally, it must be possible to detect the use of the pattern or principle using a programming-language-agnostic heuristic, because we needed to be able to detect their use in test files written in numerous languages.
After careful review, we chose four patterns that met the selection criteria. Table \ref{tbl:patternDesc} provides a brief mapping between each pattern and the quality attributes they address as well as rationale for satisfying the quality attributes based on existing works. Patterns are described in brief Coplin form~\cite{adams1996fault} in the context of test smells/quality goals below. Our detection heuristics are described in section~\ref{sec:patternFind}. 

\begin{table*}
  \caption{Overview of Testing Patterns Studied} 
  \label{tbl:patternDesc}
  \begin{center}
  \includegraphics[width=0.92\linewidth]{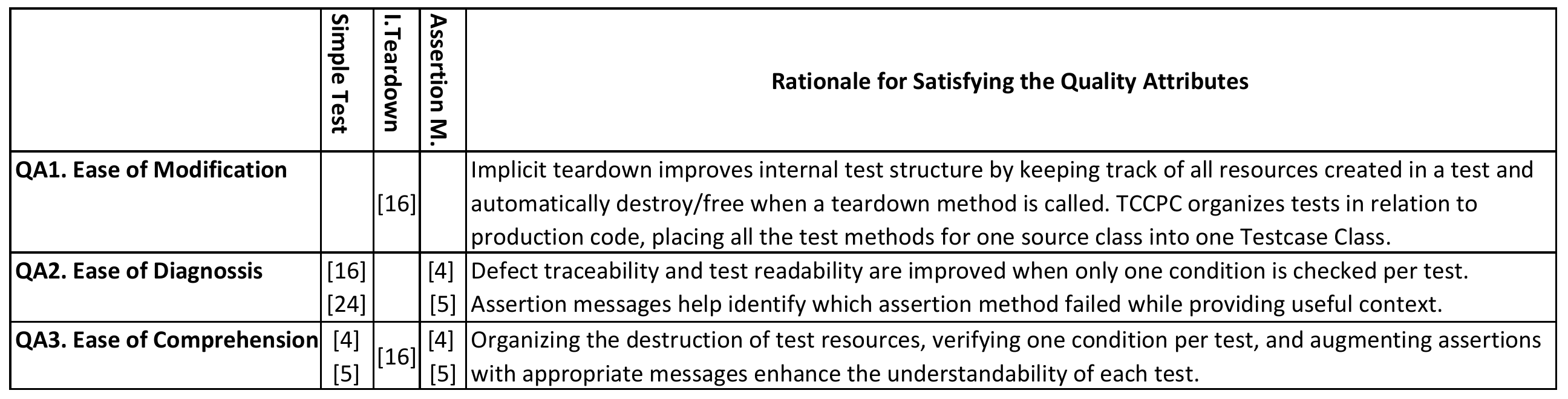}
  \end{center}
  \vspace{-15pt}
\end{table*}

\vspace{2pt}
\noindent \underline{Simple Tests}: 
\noindent\textbf{Context:}  A test with multiple assertions.

\noindent\textbf{Problem:} Obscure Test(Eager Tests)~\cite{van2001refactoring,Meszaros07} If a test verifies multiple conditions and one fails, any conditions after it will not be executed. After a failure, a test with multiple assertions is more difficult to diagnose, and any checks appearing after the failing assertion would not be executed.

\noindent\textbf{Solution:}
A \textit{simple test} is a test method which only checks one code path of the subject under test. Meszaros categorizes simple tests as a goal of test automation, which can be realized by following the \textit{Verify One Condition Per Test} principle. 

\vspace{2pt}
\noindent\underline{Implicit Teardown}:
\noindent\textbf{Context:} Objects and fixtures are created within a test class or method to simulate the right test environment. 
\noindent\textbf{Problem:} Conditional Test Logic (Complex Teardown)~\cite{Meszaros07}. 
It is difficult to ensure that test fixtures have been removed properly if the teardown code is difficult to comprehend, and leftover fixtures can affect other tests. While most objects can be automatically destroyed in languages with garbage-collection, some objects need to be destroyed manually such as connections or database records.

\noindent\textbf{Solution:}
In this pattern, a \texttt{teardown} method containing removal code for items that are not automatically removed is run after each test method execution, no matter if it passes or fails. In a class with many tests, this pattern eases maintenance by ensuring \textit{all} persisting items are removed, and promotes readability by making it easier for developers to understand which test items are being destroyed, and when.  


 \vspace{2pt}
\noindent\underline{Assertion Message}:
\noindent\textbf{Context:} Tests written with xUnit based frameworks verify conditions using \textit{Assertion Methods}. By default, these methods do not return any descriptive information when they fail. 
\noindent\textbf{Problem:} Assertion Roulette~\cite{van2001refactoring} In a test with multiple assertions, or when many tests are run simultaneously, this lack of output can make it difficult to diagnose which assertion failed.
\noindent\textbf{Solution:}
 This problem can be mitigated by adding an \textit{Assertion Message} to each assertion method. While this might seem like a trivial pattern, its presence in an assertion shows a conscious choice to improve the readability of the test. This pattern can also be used as an alternative solution to the problem discussed in \textit{Simple Test}. 


\vspace{2pt}
\noindent \underline{Testcase Class Per Class (TCCPC)}
\noindent\textbf{Context:} a project with a large number of classes.
\noindent\textbf{Problem:} Poor Test Organization. \textit{Where do I put a new test?} There are a number of options, with the worst-case being ALL project test cases in ONE test class.
\noindent\textbf{Solution:}
Developers who take this into consideration early in a project may wish to create one \textit{Testcase Class per (Source) Class} being tested. This improves readability and maintenance while making organization more straightforward. This pattern can also be applied to a mature project with existing tests, but if it contains a very large test suite, this refactoring can be time consuming.

\subsection{Testing Frameworks}
Instead of manually writing and executing complex custom tests, developers can use a unit testing framework, which provides syntax conventions and utility methods. Presence of a test framework is one criteria we use to identify test files, described in section~\ref{sec:findtests}. Mezaros'~\cite{Meszaros07} collection of test patterns are based on the \textit{xUnit} family of unit test frameworks (JUnit, CUnit, etc.). However, since there are usually multiple test frameworks for each programming language, we look for both xUnit and other frameworks. 

\section{Methodology}
\label{sec:method}
To answer our motivating question, our methodology consists of three steps: (i) retrieve a large number of open source projects from online repositories, (ii) identify test files across these projects, and (iii) detect testing patterns across test files. We discuss each of these steps in the following subsections.

The data used in this study include \totalProjects{} open source projects, \totalFrameworks{} unit testing frameworks, and 4 unit testing patterns. A full copy of this data can be found in our project website~\footnote{All the collected data and the tools developed in this paper can be found at: \url{https://goo.gl/Mc7tHk}}. To detect test files and pattern use in the projects, we developed automated, heuristic-based techniques.

\subsection{Gathering Open Source Projects}
Our first step was to extract a large-scale collection of software projects from open source repositories. Our current collection contains \totalProjects{} projects extracted from GitHub. To retrieve projects, we used GHTorrent~\cite{Gousi13}, which provides Github's public data and events in the form of MongoDB data dumps containing project metadata such as users, commit messages, programming languages, pull requests, and follower-following relations. We also used \textit{Sourcerer}~\cite{Sourcerer}, which provided versioned source code from multiple releases, documentation (if available), and a coarse-grained structural analysis.
 After extracting projects from these resources, we performed a data cleanse in which we removed all the empty projects and duplicate forks. Table \ref{tab:Repo} shows the number of projects collected per programming language.

 \begin{table}[htb]

 	\centering
 	\caption{Number of Projects included in the study} 
 	\scriptsize
  	\begin{tabular}{  p{1.4cm}c | p{1.5cm}c | p{1.8cm}c  }
 		\hline
 		Language & Freq. & Language & Freq. & Language & Freq. \\ \hline \hline
 		\texttt{JavaScript}  & 20201 & \texttt{Lua}          & 343 & \texttt{F\#}           & 21 \\ \hline
		\texttt{Java}        & 14493 & \texttt{Haskell}      & 338 & \texttt{Processing}    & 21 \\ \hline
		\texttt{Ruby}        & 8558  & \texttt{Rust}         & 241 & \texttt{Dart}          & 19 \\ \hline
		\texttt{Python}      & 7180  & \texttt{CoffeeScript} & 230 & \texttt{FORTRAN}       & 17 \\ \hline
		\texttt{PHP}         & 6889  & \texttt{Matlab}       & 214 & \texttt{Objective-C++} & 16 \\ \hline
		\texttt{C++}         & 4758  & \texttt{Groovy}       & 161 & \texttt{Haxe}          & 15 \\ \hline
		\texttt{C}           & 4369  & \texttt{Puppet}       & 99  & \texttt{Julia}         & 15 \\ \hline
		\texttt{C\#}         & 4365  & \texttt{TypeScript}   & 63  & \texttt{Elixir}        & 13 \\ \hline
		\texttt{Shell}       & 1879  & \texttt{Emacs Lisp}   & 62  & \texttt{Pascal}        & 13 \\ \hline
		\texttt{Objective-C} & 1493  & \texttt{PowerShell}   & 55  & \texttt{Prolog}        & 13 \\ \hline
		\texttt{Go}          & 858   & \texttt{Arduino}      & 46  & \texttt{Visual Basic}  & 13 \\ \hline
		\texttt{Swift}       & 782   & \texttt{ASP}          & 44  & \texttt{Common Lisp}   & 12 \\ \hline
		\texttt{Scala}       & 633   & \texttt{OCaml}        & 29  & \texttt{Mathematica}   & 11 \\ \hline
		\texttt{VimL}        & 440   & \texttt{Assembly}     & 27  & \texttt{AGS Script}    & 9  \\ \hline
		\texttt{Perl}        & 399   & \texttt{Erlang}       & 27  & \texttt{BlitzBasic}    & 9  \\ \hline
		\texttt{Clojure}     & 386   & \texttt{ActionScript} & 26  & \texttt{Scheme}        & 9  \\ \hline
		\texttt{R}           & 366   & \texttt{D}            & 23  & \texttt{Cuda}          & 8  \\ \hline
		\multicolumn{6}{l}{
        	*Total number of projects: \totalProjects    
         }
 	\end{tabular}
 	\normalsize
 	\label{tab:Repo} %
 \end{table}%

\subsection{Extracting and Analyzing Test Cases}
\label{sec:findtests}
Next, we developed a heuristical approach to identify test files among other source files and identify which automated testing frameworks were used to implement the tests. Our automated test case extraction and analysis tool contains the following components:

 \vspace{2pt}
\noindent\underline{Catalog of Automated Testing Frameworks}
This catalog contains a list of existing testing frameworks and regular expressions which represent the import statements for each framework. To establish this catalog, we started with a pre-compiled list of unit testing frameworks sorted by language from Wikipedia~\cite{wiki:frameworks}. For each framework in this list we then manually collected additional information such as applicable languages. Most importantly, we searched source code, official documentation, and tutorials to find the framework's API and code annotations in order to determine how each framework is imported into a project. With this information we developed heuristics for determining which (if any) framework a project was using. Frameworks which required payment for use, without source code or sufficient documentation, or for which we could not develop a technique to identify its import statement were then excluded from the list. Our final list contained \totalFrameworks{} testing frameworks for \totalLanguages{} different languages.

 \vspace{2pt}
\noindent\underline{Test Case Detection and Analysis}
This component used two \textit{heuristical conditions} to analyze each project in order to determine they contained test files. Our analysis used a \textit{voting mechanism} to determine if each condition had been met. The conditions were as follows: 

\vspace{3pt}
\small
\noindent \texttt{\textbf{Condition \#1} The file imported at least one automated unit testing framework}
\normalsize

For each test framework we used the list of specific import statement(s) to create regular expressions, and pattern matching was used to identify the import statement within the source code. The accuracy of this approach was evaluated using a peer review process including 50 randomly selected projects. With 100\% accuracy, this approach identifies various forms of imported testing libraries.

\vspace{3pt}
\small
\noindent \texttt{\textbf{Condition\#2} The file contained keywords commonly used by frameworks in the file’s programming language}
\normalsize
\vspace{3pt}

We also curated a list of common \textit{testing keywords} and \textit{constructs} (such as \texttt{assert} and \texttt{verify}) found in each programming language (Table \ref{tab:testKeywords}.). This list was established by manually analyzing at least 10 sample test cases per language, and collecting \textit{framework-specific terms} from the documentation of each testing framework.

We processed every file in our collection and calculated a binary value for both condition. A file received a vote of 0 for a condition if it did not meet that particular condition, or a 1 if the file did meet the condition. A file was considered a test file if both conditions received a vote of 1.

This heuristical approach to test case identification differs from most existing methods in the literature ~\cite{kochhar2013empirical,van2009establishing} such as selecting files which include the word `test' in the file name. We chose not to include this condition to avoid relying on file name conventions because we believe the use of frameworks and test-keywords are stronger indications of test files and will not exclude files that do not follow the `testX' naming convention. We also excluded other conditions such as Fixture Element Types or Static Call Graphs~\cite{van2009establishing} in order to develop language-agnostic heuristics which could be applied to all the languages in our dataset. 
\begin{table}[htbp]
	\centering
    \scriptsize
	\caption{Sample of Testing Keywords Per Language}
	\begin{tabular}{p{1cm}|p{6cm}}
		\hline
		Language& Tag Regex \\
		\hline
		Java  &  @Test, @Before, @After, @AfterClass, @BeforeClass, @Ignore, @RunWidth, @Parameters, assert.* \\\hline
		C\#softw    & [TestClass], [TestMethod], [ClassInitialize], [TestInitialize], [TestCleanUp], [TestFixture], [SetUp], [Test], [ExpectedException(typeof(.*))], [Fact], [Theory], [InlineData(.*)], [CollectionDefinition(.*)], [Collection(.*)] \\\hline
		JavaScript    & test(, describe(rb,assert(, assert\_, describe \\\hline
		Python    & def test\_, testmod( \\\hline
		C     & START\_TEST, END\_TEST, setUp(, tearDown(, UnityBegin(, RUN\_TEST(, TEST\_ASSERT\_EQUAL(, cmocka\_unit\_test(, assert.*, VERIFY( \\\hline
		C++   & START\_TEST, END\_TEST, BOOST\_CHECK(, BOOST\_REQUIRE(,  CPPUNIT\_TEST(, CPPUNIT\_ASSERT\_EQUAL(, CPPUNIT\_FAIL( \\\hline
		Perl    & ok(, is(, isnt(, like(, unlike(, cmp\_ok(, pass(, fail( \\\hline
		PHP   & TEST, FILE, EXPECT, EXPECTF, SKIPIF, INI, CLEAN, test, assert, @test \\\hline
	\end{tabular}%
	\label{tab:testKeywords}%
    \normalsize
    \vspace{-6pt}
\end{table}%

\subsection{Detecting Patterns in Test Files}
\label{sec:patternFind}
To detect the selected patterns within test files we developed another set of heuristical techniques. For each pattern, we were able to identify unique characteristics such as code structure, test method signature, and reserved keywords which indicate the presence of said pattern. 

The \textit{Implicit Teardown} and \textit{Assertion Message} patterns are built-in but \textit{optional} components of the xUnit framework, and can be best identified in source code by verifying the presence of certain keywords. The heuristical approach used to detect each pattern is described in the following:

\begin{itemize}
\item \textit{Assertion Messages} are passed as arguments into an \texttt{assert} method in a test. Common forms for an assertion with a message include: \texttt{assertCONDITION(ARG1, ARG2, [MESSAGE])}, \texttt{assert(ASSERT, CONDITION, [MESSAGE])}, and \texttt{assertBOOLEAN(ASSERT, [MESSAGE])}. By developing regular expressions for these three method signatures we can detect if an assert function contains a message. We totaled the number of matches for each signature variation in every file, and considered a test case if there was at least one match.

\item \textit{Simple Test} is detected by tracking the number of asserts in a test case. If there is only one assert, we consider it a simple test. We adopted the previous heuristic to detect this pattern.

\item  \textit{Implicit Teardown} required a semantic heuristic \textit{and} a relatively more complex search to verify its presence. Our heuristic first finds a \texttt{teardown} method, a built-in signature in testing frameworks used to implement the teardown pattern. After finding this method signature in a test method, our heuristic searches for `remove' or `destroy' methods to indicate that items are being deleted. In our heuristic, the method name `teardown' \textit{must} be found, along with at least least one of the two other methods. 

\item \textit{Testcase Class Per Class} is identified by first creating a call graph between all classes in the project, including test classes. This call graph is used to check if there is one test class which depends on a source class in production code. We used an existing commercial tool~\cite{understand} to generate the call graph. This tool can only perform static analysis on a limited number of languages. Therefore, we were only able to detect this pattern in 11 of the \totalLanguages{} languages present in our dataset: C, C++, C\#, FORTRAN, Java, Pascal, VHDL, HTML, CSS, PHP, and Javascript. 
\end{itemize}

\subsection{Validation of Data Extraction Techniques}
To validate the correctness of the data collected, we performed the following checks:
\subsubsection*{Test Files}
To verify the correctness of the test files, we randomly selected 20 projects and manually identified any test files and any test framework the projects imported. We then ran our test case analysis (conditions 1 \& 2) on these files and verified that the results matched. Our approach achieved an accuracy of 100\%.
\subsubsection*{Pattern Files}
To verify the correctness of the pattern files, we randomly selected and manually verified files which were identified by our heuristic to contain a pattern, 15 per pattern. Our pattern detection approach had 100\% accuracy.

\section{Results}
\label{sec:results}

This section presents the results for the main motivating question and the three consequent research questions.

\subsection{The State of Software Testing in Open Source}
\label{sec:teststate}
The following data provides general insights into the state of software testing in open source, specifically project size, test-to-code ratios, and framework usage across the projects studied. This provides better context and background for the interpretation of our research question results.

\noindent\underline{Distribution of Project Sizes:} As mentioned in section ~\ref{sec:projsize}, we grouped the \totalProjects{} projects in our collection into four size categories (\textit{very small}, \textit{small}, \textit{medium} and \textit{large}) to better understand how testing varies with project size. After grouping the projects, we found that our collection was 46.55\% \textit{very small} projects,  30.82\% \textit{small} projects, 16.16\% \textit{medium} projects and 6.45\% \textit{large} projects.

\noindent\underline{Test Code to Production Code Ratio:} We observed that \totalProjectsWithTests{} (17.17\%) of projects included test files. We computed the ratio of lines of code from test cases to total number of line in the production code (TPRatio~\ref{sec:metrics}) and report this metric for each project size category in Table~\ref{tab:rq1}.

The average TPRatio of all the projects in our collection was 0.06. We also note that \textit{small} and \textit{very small} projects included fewer test files. Out of 38,379 \textit{very small} projects and  25,425 \textit{small} projects, only 3,207 (8.4\%) and 4,203 (16.5\%) of projects had at least one test file, respectively. \textit{Medium} and \textit{large} projects in our dataset contained more tests (31\% and 67.2\%, respectively). 

\begin{framed}
\vspace{-3pt}
\noindent
\textbf{Testing In Open Source:}
17.17\% of projects studied contained test files. The average ratio of test code to production code was very low, 0.06. Smaller projects had a lower ratio than larger projects. One hypothesis for this result is that medium and large projects are more complex and, therefore, introduce automated testing practices more to ensure code correctness.\vspace{-3pt}
\end{framed}
\begin{table}
\centering
\caption{Test-Production Code Ratios Across Project Sizes}
\vspace{2pt}
\scriptsize
\label{tab:rq1}
\begin{tabular}{r|l|l|l|l|l|}
\cline{2-6}
\multicolumn{1}{l|}{}                             & \multicolumn{1}{c|}{\textbf{All Proj.}} & \multicolumn{1}{c|}{\textbf{Very Small}} & \multicolumn{1}{c|}{\textbf{Small}} & \multicolumn{1}{c|}{\textbf{Medium}} & \multicolumn{1}{c|}{\textbf{Large}} \\ \hline
\multicolumn{1}{|r|}{\textbf{Total \# Projects}}         & 82447                                   & 38379                                    & 25425                               & 13322                                & 5321                                \\ \hline
\multicolumn{1}{|r|}{\textbf{Avg. TPRatio}}         & 0.06                            		& 0.05                             		   & 0.06		                         & 0.08                                 & 0.08                                \\ \hline
\multicolumn{1}{|r|}{\textbf{Max. TPRatio}}         & 12.20 								& 12.20                                    & 11.05                               & 7.92                                 & 4.62                                \\ \hline
\multicolumn{1}{|r|}{\textbf{\#Proj. With Tests}} & 15119                           & 3207                              & 4203                        & 4131                         & 3578                        \\ \hline
\end{tabular}
\normalsize
\end{table}

\noindent\underline{Test Framework Adoption in the Open Source Community:} Of the \totalFrameworks{} unit testing frameworks we searched for in 48 programming languages, we detected usage of \usedFrameworks{} frameworks in 9 languages (C, C++, C\#, Java, JavaScript, Perl, PHP, Python and Ruby). Table~\ref{table:frameworks_language} presents the three most-used frameworks per language. In the case of the Perl language, we only detected the usage of two testing frameworks.

\begin{table}[h]
\centering
\caption{Top 3 Unit Testing Frameworks per Language}
\label{table:frameworks_language}
\scriptsize
\begin{tabular}{c|l|c|}
\cline{2-3}
\multicolumn{1}{l|}{}                             & \multicolumn{1}{c|}{\textbf{Framework}} & \textbf{\# Occurrences} \\ \hline
\multicolumn{1}{|c|}{\multirow{3}{*}{C}}          & Check                            		& 251                    \\ \cline{2-3} 
\multicolumn{1}{|c|}{}                            & GLib Testing                            & 98                    \\ \cline{2-3} 
\multicolumn{1}{|c|}{}                            & SimpleCTest                             & 38                     \\ \hline
\multicolumn{1}{|c|}{\multirow{3}{*}{C++}}        & Boost Test Library                      & 113                   \\ \cline{2-3} 
\multicolumn{1}{|c|}{}                            & Google Test                             & 29                    \\ \cline{2-3} 
\multicolumn{1}{|c|}{}                            & Google C++ Mocking Framework            & 29                     \\ \hline
\multicolumn{1}{|c|}{\multirow{3}{*}{C\#}}        & NUnit                                   & 443                    \\ \cline{2-3} 
\multicolumn{1}{|c|}{}                            & Moq                                     & 164                    \\ \cline{2-3} 
\multicolumn{1}{|c|}{}                            & Rhino Mocks                             & 30                     \\ \hline
\multicolumn{1}{|c|}{\multirow{3}{*}{Java}}       & JUnit                                   & 3841                   \\ \cline{2-3} 
\multicolumn{1}{|c|}{}                            & Mockito                                 & 573                    \\ \cline{2-3} 
\multicolumn{1}{|c|}{}                            & TestNG                                  & 303                    \\ \hline
\multicolumn{1}{|l|}{\multirow{3}{*}{JavaScript}} & Mocha                                   & 6714                   \\ \cline{2-3} 
\multicolumn{1}{|l|}{}                            & JSUnit                             	    & 758                   \\ \cline{2-3} 
\multicolumn{1}{|l|}{}                            & Enhance JS                              & 737                   \\ \hline
\multicolumn{1}{|c|}{\multirow{2}{*}{Perl}}   	  & Test::More                              & 49                     \\ \cline{2-3} 
\multicolumn{1}{|c|}{}                        	  & Test::Builder                           & 2                      \\  \cline{2-3} 
\multicolumn{1}{|c|}{\multirow{3}{*}{PHP}}        & PHPUnit                                 & 1000                   \\ \cline{2-3} 
\multicolumn{1}{|c|}{}                        	  & PHP Unit Testing Framework              & 867                   \\ \cline{2-3} 
\multicolumn{1}{|c|}{}                        	  & SimpleTest              				& 348                    \\ \hline
\multicolumn{1}{|c|}{\multirow{3}{*}{Python}} 	  & unittest                                & 1663                   \\ \cline{2-3} 
\multicolumn{1}{|c|}{}                        	  & Autotest                                & 466                    \\ \cline{2-3} 
\multicolumn{1}{|c|}{}                        	  & Nose                                    & 336                    \\ \hline
\multicolumn{1}{|c|}{\multirow{3}{*}{Ruby}}   	  & Test::Unit                              & 202                   \\ \cline{2-3} 
\multicolumn{1}{|c|}{}                        	  & Shoulda                                 & 75                    \\ \cline{2-3} 
\multicolumn{1}{|c|}{}                        	  & minitest                                & 56                    \\ \hline
\end{tabular}
\normalsize
\end{table}

Relying only on the total number of projects that adopted a framework to identify the most commonly used ones would result in a bias. As presented in Table~\ref{tab:Repo}, the distribution of the programming languages over the projects in our repository is not uniform. Therefore, the frameworks that are used in the most frequent languages in our repository would more likely be placed at the top of the ranking. To overcome this, we normalized the frequency of the frameworks by dividing the \emph{framework frequency} by the \emph{total number of projects in the framework's programming language(s)}. Table~\ref{table:top_frameworks} enumerates the ten most used frameworks sorted by this normalization value. From this table we observe that \texttt{Mocha}, a testing framework for JavaScript projects, is the most used framework.

\begin{table}[h]
\centering
\caption{Top 20 Most Used Frameworks}
\label{table:top_frameworks}
\scriptsize
\begin{tabular}{|l|c|c|c|}
\hline

\textbf{Framework} 			& \textbf{Language} & \textbf{Freq.} & \textbf{Normalized Freq.}		\\ \hline
Mocha						&	JavaScript		&	6714		&	0.33							\\ \hline
JUnit						&	Java			&	3841		&	0.27							\\ \hline
unittest					&	Python			&	1663		&	0.23							\\ \hline
PHPUnit						&	PHP				&	1000		&	0.14							\\ \hline
PHP Unit Testing	&	PHP				&	867			&	0.13							\\ \hline
Test::More					&	Perl			&	49			&	0.12							\\ \hline
NUnit						&	C\#				&	443			&	0.10							\\ \hline
Autotest					&	Python			&	466			&	0.06							\\ \hline
Check						&	C				&	251			&	0.06							\\ \hline
SimpleTest					&	PHP				&	348			&	0.05							\\ \hline
Nose						&	Python			&	336			&	0.05							\\ \hline
Mockito						&	Java			&	573			&	0.04							\\ \hline
Moq							&	C\#				&	164			&	0.04							\\ \hline
JSUnit						&	JavaScript		&	758			&	0.04							\\ \hline
Enhance JS					&	JavaScript		&	737			&	0.04							\\ \hline
Sinon.js					&	JavaScript		&	638			&	0.03							\\ \hline
py.test						&	Python			&	213			&	0.03							\\ \hline
Chai						&	JavaScript		&	523			&	0.03							\\ \hline
Vows						&	JavaScript		&	517			&	0.03							\\ \hline
Boost Test Library			&	C++				&	113			&	0.02							\\ \hline
\end{tabular}
\begin{flushleft}
Normalized Freq= Framework Freq. Divided by the Total \#Projects in that 
Programming Language\\ 
\end{flushleft}
\normalsize
\end{table}

\begin{framed}
\vspace{-3pt}
\noindent
\textbf{Test Framework Usage:} 
From 251 testing frameworks included in our study, only 93 were imported by developers in open source projects. Mocha (Javascript) was the most commonly used framework. 
\vspace{-3pt}
\end{framed}
\subsection{Satisfaction of Quality Attributes through Pattern Adoption}
\label{sec:discussions}

We identified \totalPatternProjects{} projects that contain at least one pattern (23.76\%) and \totalAllPatternProjects{} projects which contained all four patterns (0.08\%). 
Table ~\ref{tab:patternFiles} shows the number of test files and projects which implement each pattern. \textit{Assertion Message} was the most commonly adopted pattern, occurring in 49,552 test files among the 399,940 test files found in projects in our collection. The other patterns were not as widely adopted. \textit{Testcase Class Per Class} was adopted in 4,565 test files, \textit{Simple Test} was only applied in 1,614 test files, and \textit{Implicit Teardown} was used in 920 test files. Therefore, few test cases applied patterns which can impact maintainability attributes. All further analysis focuses on projects with tests containing patterns. 

\begin{table}[h]
\centering
\caption{Pattern Detection Results}
\scriptsize
\label{tab:patternFiles}
\begin{tabular}{|l|c|c|}
\hline
\textbf{Pattern} 	& \textbf{\# Files}  & \textbf{\# Projects}\\ \hline
Assertion Message	&	49552 & 3048 \\ \hline
Simple Test			&	1614 & 531    \\ \hline
Implicit Teardown	&	920 & 153    \\ \hline
TCCPC &	4565 & 810	 \\ \hline
\end{tabular}
\normalsize
\end{table}

\begin{figure}[!ht]
  \caption{Pattern Adoption Ratios (Excludes No-Pattern Projects)}\label{fig:boxplot}
  \centering
    \includegraphics[width=0.44\textwidth]{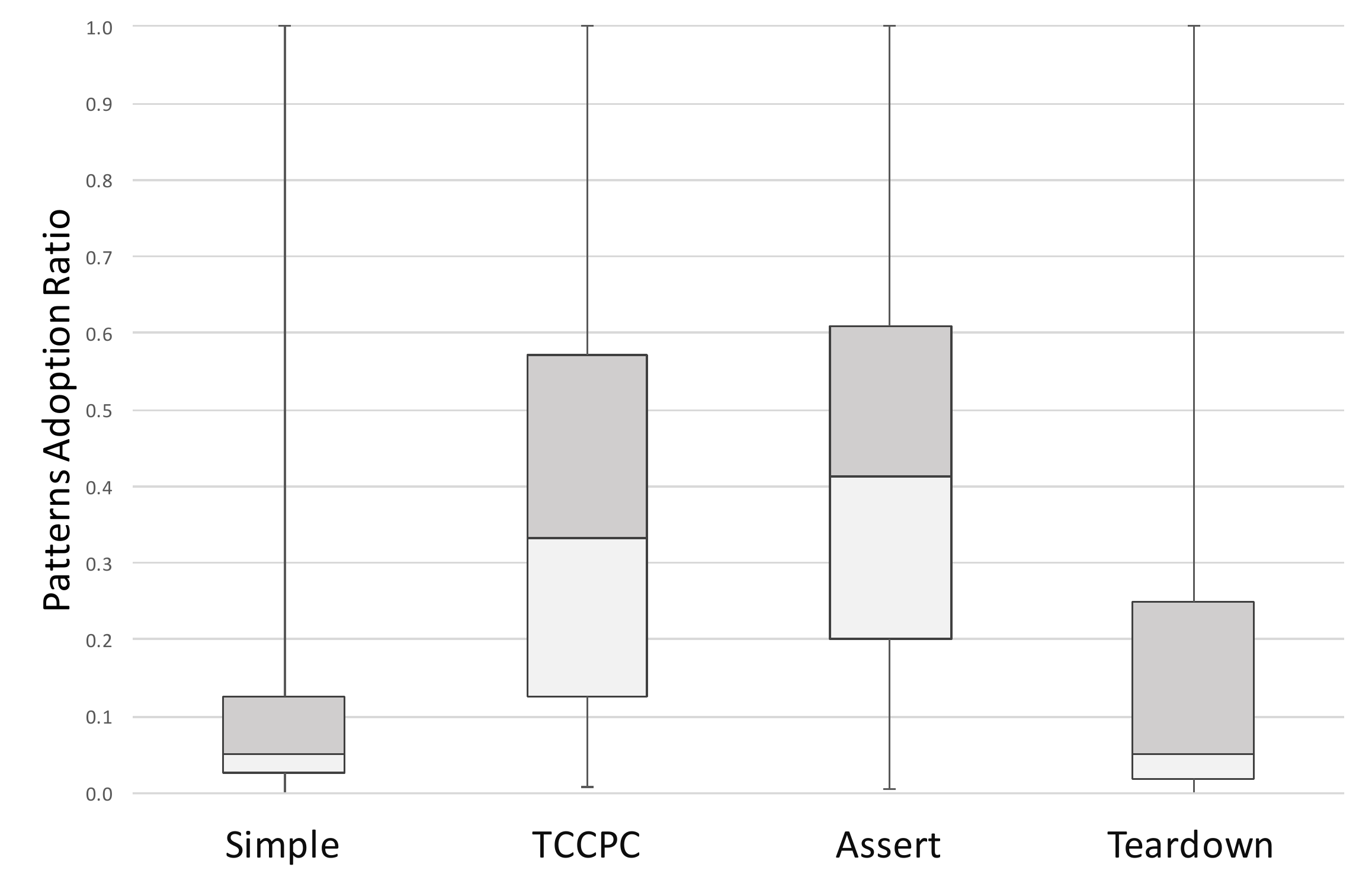}
\end{figure}

\noindent\underline{Pattern Adoption Ratio:}
In order to better understand \textit{how often} the patterns were adopted in projects, we computed the adoption ratio of the patterns as the \emph{number of test case files with the pattern} divided by the \emph{total number of test files} in the project. This ratio was computed per pattern for the projects that adopted each pattern.  The adoption ratios of patterns are presented in Figure~\ref{fig:boxplot}. From this figure, we find that both \emph{Assertion Message} and \emph{TCCPC} had the highest adoption rates, which means that these patterns were consistently applied throughout the projects' test cases. In fact, the median adoption rates for \emph{Assertion Message} and \emph{TCCPC} were 41.2\% and 33.3\%, respectively. 
Conversely, the adoption ratios for \emph{Simple Test} and \emph{Implicit Teardown} were both 5\%.

\noindent\underline{Pattern Adoption And Number of Test Cases:} In addition to the analysis of the raw results, we wanted to examine if patterns are typically found in projects containing a lot of test cases or not. Therefore, we built a linear regression model to examine the relationship between the adoption ratio of each pattern (independent variables) to the number of test cases (dependent variable) in the corresponding project. However, since the number of test cases could just be a function of the size of a project, we include the size of the project (in KLOC) as a control variable in the regression model. 

From Table~\ref{tab:patternlinReg}, we find that as expected there is a strong positive relationship between size of the project and the number of test cases. However, surprisingly, we notice that the relationship between adoption ratio and the number of test cases is negative. This implies that smaller projects are more disciplined in using test case patterns. 
\begin{table}[h]
\centering
\caption{Linear Regression Coefficients Projects}
\scriptsize
\label{tab:patternlinReg}
\begin{tabular}{p{1.9cm}|c|c|p{0.708cm}|c}
\hline
                  & Estimate   & Std. Error&t value & $Pr(>|t|)$ \\ \hline
(Intercept)       & 7.285e+01 & 4.780e+00 & 15.243 & $<$ 2e-16   \\ \hline
Simple Test       &-1.373e+02 & 3.191e+01 & -4.303 & 1.73e-05    \\ \hline
TCCPC         	  &-7.658e+01 & 1.104e+01 & -6.938 & 4.71e-12    \\ \hline
Assertion Message &-5.872e+01 & 8.391e+00 & -6.998 & 3.09e-12    \\ \hline
Teardown          &-9.289e+01 & 4.200e+01 & -2.212 & 0.027       \\ \hline
TotalLinesOfCode  & 2.079e-05 & 1.754e-06 & 11.849 & $<$ 2e-16   \\ \hline
\end{tabular}
\end{table}
\begin{framed}
\vspace{-3pt}
\noindent
\textbf{Pattern Adoption:}
Open source projects do not frequently adopt the studied testing patterns. Smaller projects have fewer test files and test-to-production code ratios, but a higher frequency of pattern adoption compared to larger projects.
\vspace{-3pt}
\end{framed}

To answer our three research questions, we used a QA-Pattern mapping (See Table ~\ref{tbl:patternDesc}) to relate the patterns we chose to study with the quality attributes they best represent. From the \totalPatternProjects{} projects which included patterns, for each quality attribute we calculated the percentage of projects that implemented each mapped pattern as well as all mapped patterns. This data is shown in Table ~\ref{tbl:QA}. Note that percentages are based \textit{only} on the total number of projects with patterns.

\begin{table*}
  \caption{Percentage of Projects Satisfying Each Maintainability Quality Attributes Using Testing Patterns}
  \label{tbl:QA}
  \begin{center}
  \includegraphics[width=0.99\linewidth]{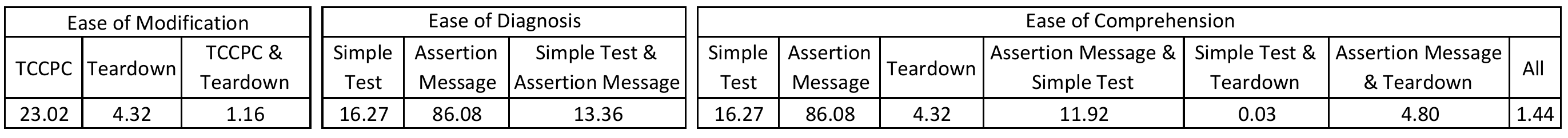}
  \end{center}
  \vspace{-15pt}
\end{table*}

\subsection*{\textbf{RQ1:} What percentage of projects with patterns addressed the \textit{Ease of Modification} Quality Attribute?}

The testing patterns which can satisfy the \textit{Ease of Modification} quality attribute for tests are \textit{Implicit Teardown} and \textit{Testcase Class Per Class}. 143 (4.32\%) projects applied \textit{Implicit Teardown}, improving modifiability of internal test contents. 815 (23.02\%) projects implemented \textit{Testcase Class Per Class (TCCPC)}, improving the modifiability of test structure in relation to production code. 41 (1.16\%) of projects implemented both patterns, addressing both internal and structural test modifiability. 
\begin{framed}
\noindent
\textbf{Ease of Modification:}
\vspace{-3pt}
Approximately one-fourth of the projects with patterns implemented the \textit{Testcase Class per Class} pattern, providing a maintainable design for test classes and their associated production classes. Fewer projects adopted \textit{Implicit Teardown}, which helps structure the destruction of test fixtures.
\vspace{-3pt}
\end{framed}

\subsection*{\textbf{RQ2:} What percentage of projects with patterns addressed the \textit{Ease of Diagnosis} Quality Attribute?}
To satisfy \textit{Ease of Diagnoses}, we looked for adoption of \textit{Simple Test} and \textit{Assertion Message} patterns. Both patterns improve defect traceability and test readability. From all projects that implement patterns, 576 projects (16.27\%) implemented \textit{Simple Test}, and 3,048 projects (86.02\%) implemented \textit{Assertion Message}. 473 (13.36\%) projects implemented both, strongly improving defect traceability and test readability. Although these two patterns are intuitive to enforce, a large number of projects have not implemented them.
\begin{framed}
\vspace{-3pt}
\noindent
\textbf{Ease of Diagnoses:}
Despite benefits in diagnoses of the \textit{Simple Test} and \textit{Assertion Message} patterns, less than one fourth of projects adopted both or \textit{Simple Test} without \textit{Assertion Message}. However, over three fourths of projects used \textit{Assertion Message} without \textit{Simple Test}.
\vspace{-3pt}
\end{framed}
\subsection*{\textbf{RQ3:} What percentage of projects with patterns addressed \textit{Ease of Comprehension} Quality Attribute?}
Comprehension improvement involves a combination of test modifiability and diagnosis improvements. The testing patterns which can satisfy \textit{Ease of Comprehension} in our study were \textit{Implicit Teardown}, \textit{Assertion Message}, and \textit{Simple Test}.
As individual usage of these patterns has been reported above, we will now discuss combinations of these patterns. 422 (11.92\%) projects implemented \textit{Assertion Message} and \textit{Simple Test}, 170 (4.80\%) projects implemented \textit{Assertion Message} and \textit{Implicit Teardown}, and 1 (0.03\%) project implemented \textit{Simple Test} and \textit{Implicit Teardown}. 51 (1.44\%) implemented all three patterns.
\begin{framed}
\vspace{-3pt}
\noindent
\textbf{Ease of Comprehension:}
Comprehension can be addressed with three of the patterns studied, but less than 2\% of projects with patterns implemented all three. The most implemented combination was \textit{Assertion Message} and \textit{Simple Test} (11.92\%), which are more intuitive to implement than \textit{Implicit Teardown}.
\vspace{-3pt}
\end{framed}

From the results reported for our three RQs, we can see that the quality attribute addressed in the highest number of projects is \textit{Ease of Diagnoses}. This indicates that developers most often applied the patterns which could help them identify why a test has failed. Relatively, \textit{Ease of Modification} and \textit{Ease of Comprehension} were not satisfied across many open source projects.
To further investigate these findings, we report three qualitative analyses on projects with different pattern usage in the following section.
\vspace{-5px}
\section{Qualitative Analyses}
\label{sec:patternanalysis}
In addition to the quantitative results reported in the previous section, we have conducted three qualitative studies to understand differences between projects with different levels of pattern adoption. First we investigated whether there are any additional project characteristics which influenced pattern adoption. Next, we studied the contributions of developers to the test cases with pattern within projects which applied all four patterns. Finally, we examined test files from projects with no patterns to determine if they used other approaches to address test maintainability.

\subsection*{Analysis \#1: Pattern Adoption and Project Characteristics}
To identify potential correlations between test pattern adoption and other project characteristics, we collected project artifacts such as source code, online repositories, and documentation for 29 randomly selected projects from 3 categories: 9 projects that applied all 4 patterns, 10 projects that applied 1, 2, or 3 (\textit{Some}) patterns, and 10 that applied 0 (\textit{No}) patterns. We looked at test documentation, organization, `coverage', industry sponsorship, and number of contributors per file.
\noindent\underline{Project Selection:}
All projects in our dataset were sorted by randomly generated numbers. 10 projects from each category (\textit{All}, \textit{Some}, and \textit{No} patterns) were then randomly chosen from the sorted list. To ensure that projects were suitable for our study, we developed exclusion criteria. Selected projects were verified that: (1) there was at least one online artifact available such as a Github page, (2) contributor information was available, and (3) the project contained original source code. For projects with \textit{Some} patterns, we also tried to choose projects with different combinations of patterns. If a project failed to meet these criteria, it was excluded from the study and a new project was randomly selected using the process above. Of the 11 total projects in our dataset which applied all 4 patterns, 2 had to be excluded, so we were only able to study 9. 

\noindent\underline{Analysis Results:}

\noindent\textbf{Project Demographics:}
For all 29 projects selected, the number of contributors ranges from 2 to 620, and the number of forks ranges from 0 to 3,108. 16 unit testing frameworks were used, and projects were written in Java, Javascript, C\#, C, C++, Python, Scala, and PHP.

\noindent\textbf{Test Documentation \& Rules:} Our first question was whether the presence of test-specific documentation affected pattern adoption. Therefore, we manually searched available artifacts to see if projects had any guidelines (documentation) or templates specifically for unit testing code. We also wanted to know if projects had rules for new contributions, specifically that existing tests must pass, and new code requires new unit tests.

Projects with \textit{All} patterns required the developers to add new tests alongside source code contributions more frequently than the projects with \textit{Some} or \textit{No} patterns. Projects with \textit{Some} patterns most often required that existing tests run and pass (6) and also included more guidelines for writing tests (4). We found instances of unit test template code in 2 projects, one with \textit{All} patterns and the other with \textit{No} patterns.

\noindent\textbf{Test Organization \& `Coverage':}
We examined if projects with well-organized tests were more likely to adopt patterns. To measure how well tests were organized in a project, one author searched its directories to find where all test files were located. Project test organization was ranked \textit{Great}, \textit{Good}, \textit{OK}, or \textit{Poor} based on how easily \underline{all} tests in the project could be located and if they were organized by component. Due to the range of languages used, we were unable to quantitatively measure coverage, so we considered `coverage' in terms of how many project components had \textit{any} tests. For test organization, only 1 project (\textit{All}-pattern) received a \textit{Great} rating, but projects with \textit{Some} and \textit{No} patterns had equal amounts of \textit{Great}(0), \textit{Good}(7), and \textit{Poor}(1) ratings. Surprisingly, projects with \textit{All} patterns had the highest amount of \textit{Poor}ly organized projects (2). For `coverage', more projects with \textit{Some} patterns had tests for all components (7), and projects with \textit{No} tested more than half of components (4).

\noindent\textbf{Industry Contribution:}
Company sponsorship of a project is another characteristic which could potentially affect pattern adoption. We also noted if the `core contributors' of a project (identified through acknowledgments and Github metrics) worked for the sponsoring company.  Of the 12 projects with industry sponsors, 4 applied \textit{All} patterns, 2 applied \textit{Some} patterns, and 6 applied \textit{No} patterns.
8 projects had rules for unit testing (2 \textit{All}, 2 \textit{Some}, and 4 \textit{No}) and 6 included test specific guidelines or templates (1 \textit{All}, 1 \textit{Some}, and 4 \textit{No}). 2 projects had \textit{Poor} test organization (1 \textit{All}, 1 \textit{No}), 9 had \textit{Great} or \textit{Good} organization (4 \textit{All}, 1 \textit{Some}, 4 \textit{No}), and 1 (\textit{No} patterns) had \textit{OK} organization. The majority of industry projects had high `coverage', with only 1 (\textit{Some}) project testing less than half of its components.

\noindent\textbf{\# Contributors Per Test File:}
Finally, we were interested in a possible correlation between pattern adoption and the number of contributors who work on an individual test file. Across all projects, the number of contributors to a single test file ranged from 1 to 10. Projects with \textit{All} or \textit{Some} pattern usage did not have higher contributors per test file than projects with no pattern usage. The average for all projects was 1 to 2 contributors per file.

\noindent\underline{Analysis Conclusions:}
The only characteristic found most frequently in projects with \textit{All} patterns was requiring new tests to be added to contributions, and one instance of \textit{Great} test organization. Projects with \textit{Some} and \textit{No} patterns were more similar than projects with \textit{All} and \textit{Some} patterns in their inclusion of these characteristics as well. Interestingly, the differences in occurrences of these characteristics across all project categories was very small. It does appear, however, that projects with industry sponsors often addressed test organization, testing rules \& guidelines, and coverage. However, only 6 of these projects implemented \textit{All} or \textit{Some} patterns. Because we were unable to identify a strong influence on pattern adoption in this study, we performed another qualitative analysis on the contributors to projects using \textit{All} patterns.
\subsection*{Analysis \#2: Contributors to Projects Using All Patterns}
Next, we investigated the influence of developers on the adoption of testing patterns. For this analysis, we identified the developers that contributed to test files in the 9 projects with \textit{All} patterns. To do this, we retrieved the commit logs of these projects and identified the developers that committed code to each test file with a pattern and computed the \textit{total} number of testing files with patterns that \textit{each} developer contributed to.

\begin{figure}[!ht]
\centering
\captionsetup{justification=centering}
 \caption{Proportion of Contributions to Test Files with Patterns Across the Investigated Projects that Implemented All Patterns (Each Color is a Contributor)}\label{fig:contributions}
    \includegraphics[width=0.49\textwidth]{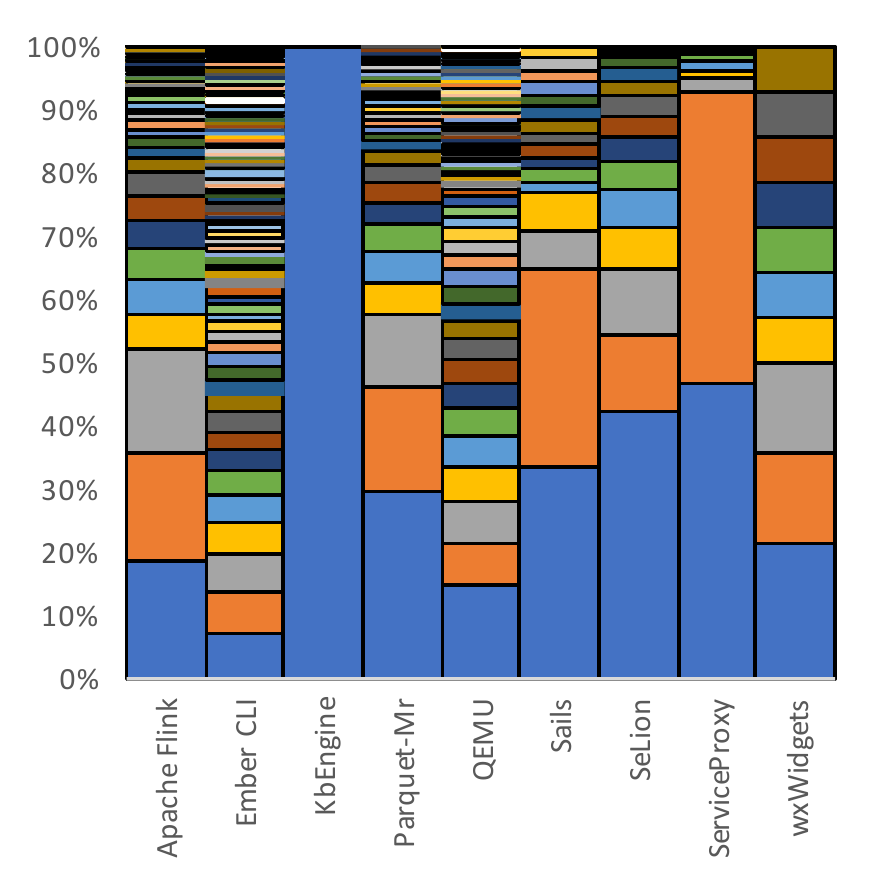}
    \normalsize
\end{figure}

\noindent\underline{Analysis Results:} 
By observing the total number of test pattern files each developer contributed to, we found that a \textit{fewer} number of developers were contributing to \textit{most} of the testing files with patterns. 
Figure~\mbox{\ref{fig:contributions}} shows the proportion of contributions to testing files with patterns made by each developer in the projects. In this figure, each color represents one developer that contributed to a testing pattern within the project. From this, we observe that in almost all projects between 2 to 3 developers are contributing the most to test files with patterns. For KbEngine, there was only one developer that was contributing to testing patterns, who is also the creator and maintainer of the project. For Ember Cli, only 9 of 100 total developers who wrote test cases were contributing most of the test files.

\noindent\underline{Analysis Conclusions:} We observed that fewer developers were contributing to the majority of test files with patterns. This suggests that testing patterns adoption in a project is an \textit{ad-hoc} personal decision made by individual developers rather than a project-wide effort to improve test quality.

\subsection*{Analysis \#3: Tests in Projects Using No Patterns}
Our final analysis was an investigation into test cases from projects which did not use patterns. Since these files contain no patterns, our goal was to see what, if any, other techniques were used to satisfy \textit{ease of modifiability, diagnosis, and comprehension}. 5 test files from each of the 10 projects without patterns used in the first qualitative study were randomly selected (50 total files) containing a total of 219 test methods. Our search was focused on solutions which implemented test failure feedback (\textit{Assertion Message}) for ease of diagnoses and fixture destruction (\textit{Implicit Teardown}) for ease of modifiability, but we also looked for characteristics which would improve comprehension of the tests such as naming conventions and the presence of comments.

\noindent\underline{Analysis Results:} 
The first general observation from this study was that within a project, files were usually consistent in their use of naming conventions and comment usage, which helped with \textit{comprehension} of the test cases. This was not the case in only 3 projects.  Only 34\% of files used comments to describe the behavior or purpose of a unit test. 54\% of projects used clear, natural language naming conventions for file, method and variable names. 38\% of files (written in Javascript) used an alternative naming convention called \texttt{describe}, a special method where the functionality of the test is passed as a parameter. The remaining 8\% of files used sequential numbering. 

As an alternative to \textit{Implicit Teardown}, 74\% of test files relied solely on garbage collection and the remaining 26\% of projects used custom methods (such as \texttt{cleanup}). 82\% of the files used an alternative to \textit{Assertion Message}: Most Javascript files studied used an alternative assertion library called `Should', which uses an \texttt{it('should...')} method syntax rather than an assertion (38\%). On failure, this message is displayed. 26\% of files used exception throwing for errors, 6\% used try/catch blocks, 4\% used custom logging objects, and the remaining 26\% used no failure feedback at all.

\noindent\underline{Analysis Conclusions:}
From these observations it is clear that the projects with no patterns use other techniques besides the XUnit patterns to address ease of maintainability and diagnoses, such as readable naming conventions and alternative failure feedback techniques. However, the data also shows that these projects seldom use comments to describe test behavior, which would improve comprehension. A future study to investigate these alternative techniques may reveal interesting results.
\begin{framed}
\vspace{-3pt}
\noindent\textbf{Qualitative Conclusions:}
Pattern adoption is independent of project characteristics, and it is dependent on individual developers. Project without patterns instead used existing libraries and proprietary and primitive techniques to satisfy quality attributes.
\vspace{-3pt}
\end{framed}
\section{Threats to Validity}
\label{sec:threats}
The following threats to the validity are identified.

\subsection{Construct Validity}
We assumed that all test cases written for a project were included in the project's source code repository. Thus, if zero test files were detected, we say the project does not contain tests.
Our approach detected the \textit{presence} of test files in the project but not how or if they were run, so we do not know if the tests are still relevant. However, this is of low concern because we are more interested in the \textit{quality} of the tests present in the project. 

\subsection{Internal Validity}
When mining software repositories, it is important to consider that some repositories may not contain projects, or may be a `fork' of another project. To prevent repeated data in our results, we removed all empty and duplicate project forks. For duplicates, we removed all but the original project \textit{or} the oldest fork if we did not have the original project in our dataset.

\subsection{External Validity}
The \totalProjects{} projects studied do not provide a complete perspective of the entire open source community. Projects may have custom tests, or use a testing framework which which was not included in this study. However, the large volume of projects and frameworks included allow new insights into the current state and quality of open source testing. Also, while the use of only four testing patterns was investigated, we believe their use is a good indication that developers took quality into consideration when designing their tests. We do \textbf{not} claim that a project without the patterns studied does not address quality as it is possible that other patterns or techniques were applied.

\subsection{Reliability Validity}
We used heuristical methods (defined in section~\ref{sec:method}) to detect test files and patterns. This approach is similar to those used in other studies (see section~\ref{sec:related}), but there are related risks. It is possible that some test frameworks were not included in our detection process because there are multiple ways to import a framework into a project. To reduce this risk, we maually searched all available documentation for each framework to find import statements. We also used a list of reserved testing keywords in our detection tool. Further, while we searched for all possible ways of implementing the patterns within the frameworks considered, it is possible that we missed instances of pattern use if the implementation did not match our heuristics.
Finally, the static analysis tool~\cite{understand} used to generate the file dependency data for detecting the Testcase Class Per Class pattern does not recognize all of the languages in our dataset. Therefore, our results for this pattern only included the projects which we were compatible with Understand. In the future we would like to detect use of this pattern in all the languages in our dataset.

\section{Conclusion}
In this paper we have performed a large-scale empirical analysis on unit tests in \totalProjects{} open source software projects written in \totalLanguages{} languages. 
We found \totalProjectsWithTests{} (17\%) projects containing test files written in 9 languages using \usedFrameworks{} test frameworks. \totalPatternProjects{} (24\%) projects applied at least one of the 4 patterns, and \totalAllPatternProjects{} projects applied all 4 patterns. 
\textit{Ease of Diagnoses} was the most commonly addressed quality attribute, and \textit{Assertion Message} was the most adopted pattern. We also found that although smaller projects contain fewer test files than larger projects, they apply patterns more frequently. 
Through three qualitative analyses of the projects with and without patterns, we found that pattern adoption is an ad-hoc decision by individual project contributors.  In summary we find that open source projects often do not adopt testing patterns that can help with maintainability attributes. More research is needed to understand why and how we can help developers write better test cases that can be maintained and evolved easily.

\section*{Acknowledgments}
This work was partially funded by the US National Science Foundation under grant numbers CCF-1543176.
\bibliographystyle{abbrv}
\bibliography{ICST}

\end{document}